\documentclass[twocolumn,superscriptaddress,prb]{revtex4-2}

\usepackage{dcolumn}
\usepackage{bm}
\usepackage{subfigure}
\usepackage[normalem]{ulem}
\usepackage{enumerate}
\usepackage{cancel}
\usepackage{flushend}
\usepackage{bbold}
\usepackage{mathtools}
\usepackage{float}
\usepackage{stmaryrd}
\usepackage{colortbl}
\usepackage[table]{xcolor}
\usepackage{array,ragged2e}
\usepackage{amssymb}
\usepackage{wasysym,graphicx,multirow,textcomp}
\usepackage{url}
\usepackage{romannum}

\usepackage{changes}

\usepackage[colorlinks = true,linkcolor = red,citecolor = magenta]{hyperref}
\usepackage[sort&compress]{natbib}

\usepackage{tikz-cd}

\makeatletter
\newcommand{\thickhline}{%
    \noalign {\ifnum 0=`}\fi \hrule height 1pt
    \futurelet \reserved@a \@xhline
}
\newcolumntype{"}{@{\hskip\tabcolsep\vrule width 1pt\hskip\tabcolsep}}
\makeatother

\newcommand{\figsizeone}{1.0}

\begin{document}

\draft

\title{Pseudo-Hermitian Topology of Multiband Non-Hermitian Systems}

\author{Jung-Wan Ryu}
    \altaffiliation{These authors contributed equally.}
    \address{Center for Theoretical Physics of Complex Systems, Institute for Basic Science (IBS), Daejeon 34126, Republic of Korea}
    \address{Basic Science Program, Korea University of Science and Technology (UST), Daejeon 34113, Republic of Korea}

\author{Jae-Ho Han}
    \altaffiliation{These authors contributed equally.}
    \address{Department of Physics, Korea Advanced Institute of Science and Technology (KAIST), Daejeon 34141, Republic of Korea}
    
\author{Chang-Hwan Yi}
    \address{Center for Theoretical Physics of Complex Systems, Institute for Basic Science (IBS), Daejeon 34126, Republic of Korea}

\author{Hee Chul Park}
    \email{hc2725@gmail.com}
    \address{Department of Physics, Pukyong National University, Busan 48513, Republic of Korea}

\author{Moon Jip Park}
    \email{moonjippark@hanyang.ac.kr}
    \address{Department of Physics, Hanyang University, Seoul 04763, Republic of Korea}

\date{\today}

\begin{abstract}
The complex eigenenergies and non-orthogonal eigenstates of non-Hermitian systems exhibit unique topological phenomena that cannot appear in Hermitian systems. Representative examples are the non-Hermitian skin effect and exceptional points. In a two-dimensional parameter space, topological classifications of non-separable bands in multiband non-Hermitian systems can be established by invoking a permutation group, where the product of the permutation represents state exchange due to exceptional points in the space. We unveil in this work the role of pseudo-Hermitian lines in non-Hermitian topology for multiple bands. In particular, the non-separability of non-Hermitian multibands can be topologically non-trivial without exceptional points in two-dimensional space. As a physical illustration of the role of pseudo-Hermitian lines, we examine a multiband structure of a photonic crystal system with lossy materials. Our work builds on the fundamental and comprehensive understanding of non-Hermitian multiband systems and also offers versatile applications and realizations of non-Hermitian systems without the need to consider exceptional points.
\end{abstract}


\flushbottom
\maketitle

\section{Introduction} Gain and loss in open quantum systems are the producers of lifetime imbalance between different eigenstates. When lifetime asymmetry manifests the decay or growth of a state, instability appears in the long-time dynamics. On the other hand, it has been shown that if a non-Hermitian system poses additional parity--time ($\mathcal{P}\mathcal{T}$) symmetry, the eigenenergies are real-valued, physically manifesting a symmetry-protected lifetime balancing of the dynamical system \cite{Ben98, ElG07, Mus08, Mak08, Kla08, Guo09, Rue10, Jog10, Sch11, Reg12, Laz13}. The dynamical stability of the eigenstates is in this way restored. Mathematically, it has been shown that $\mathcal{P}\mathcal{T}$ symmetry can be understood as a pseudo-Hermiticity that ensures the appearance of complex-conjugate-paired eigenenergies in a more general setting \cite{Mos02a, Mos02b}.

Complex-valued energy harbors an intriguing topological structure known as an exceptional point (EP), which is characterized by the vorticity of the eigenenergy along a loop that encircles the EP \cite{Kat66, Hei90, Hei04, Dem04, Gao15, Dop16, She18, Yoo18, Zha19, Che20, Tan20, Li20, Liu20, Yan21, Yu21, Shu22, Sch22}. EPs and their variations such as exceptional lines, exceptional rings, and Hopf bundles in higher-dimensional systems have been used to describe the topological structures of multibands \cite{Xu17, Car18, Cer19, Wan19, Car19, Yan19, Xia20, Zha20, He20, Ber21, Zha21, Wan21a, Liu22, Guo23, Zha23, Pak23}. In two-dimensional (2D) parameter space, the points satisfying the pseudo-Hermiticity condition form lines and are usually accompanied by EPs. Two kinds of PHL can result: imaginary and real. On an imaginary PHL where the imaginary parts of the eigenenergies coincide, the system is in a $\mathcal{PT}$-symmetric phase that has two modes with different frequencies but the same decay rates. This represents dynamical stability between the eigenstates. For instance, in the Brillouin zone of a lattice model, imaginary PHLs represent momentum lines that have long-lifetime excitation. On the other hand, the Hamiltonian on a real PHL represents $\mathcal{PT}$-broken phases having two conjugate eigenenergies with the same frequency but different decay rates. Mathematically, the Hamiltonian on imaginary and real PHLs can be transformed into Hermitian and anti-Hermitian Hamiltonians.

Recent studies on the classifications of non-Hermitian topological phases can be categorized into three groups: those using (i) complex Berry phases \cite{Gar88, Dat90, Mos99, Lia13}, (ii) winding numbers of complex energy bands \cite{Yao18a, Gon18, Lee19, Gha19, Wan21c, Din22}, and (iii) homotopy equivalence classes for non-separable bands \cite{Ryu12, Zho18, Pap18, Li21, Woj20, Hu21, Wan21b, Pat22, Koe23, Zho23, Ryu23}. For a homotopy classification, one considers a one-dimensional periodic loop in the 2D parameter space and the resultant eigenstate exchange effect. State exchange occurs during encircling EPs, and the eigenstate exchange effect can be described by the equivalence classes of the permutation group that characterizes the EPs \cite{Ryu22, Ryu23, Zho23}. The classes are non-trivial if state exchange occurs; otherwise they are trivial. We define the 2D bands as non-separable bands if there is at least a one-dimensional periodic loop on which the initial state exchange into the different state after a period on the 2D space. Though the homotopy classification is a general issue independent of EPs, the classification has been intensively studied in non-Hermitian systems with EPs because of the intrinsic non-trivial topology of EPs. While many previous works have focused on the eigenstate switching effect in the presence of EPs, in this work, we show that this effect can exist only with PHLs, without EPs, in a given 2D space and thus reveal the origins. That is, there exist non-trivial classes of permutation groups in a parameter space without EPs, due to the topology of a parameter space. In addition, we show that the classes undergo a phase transition through the creation and annihilation of EPs. We also study the stability of PHLs in multiband systems with EPs to show that the topology of the parameter space as well as EPs originate the stability.

\section{PHLs and topology of parameter space}

\subsection{PHLs in non-Hermitian systems}
\begin{figure}
\centering
\includegraphics[width=1.0\linewidth]{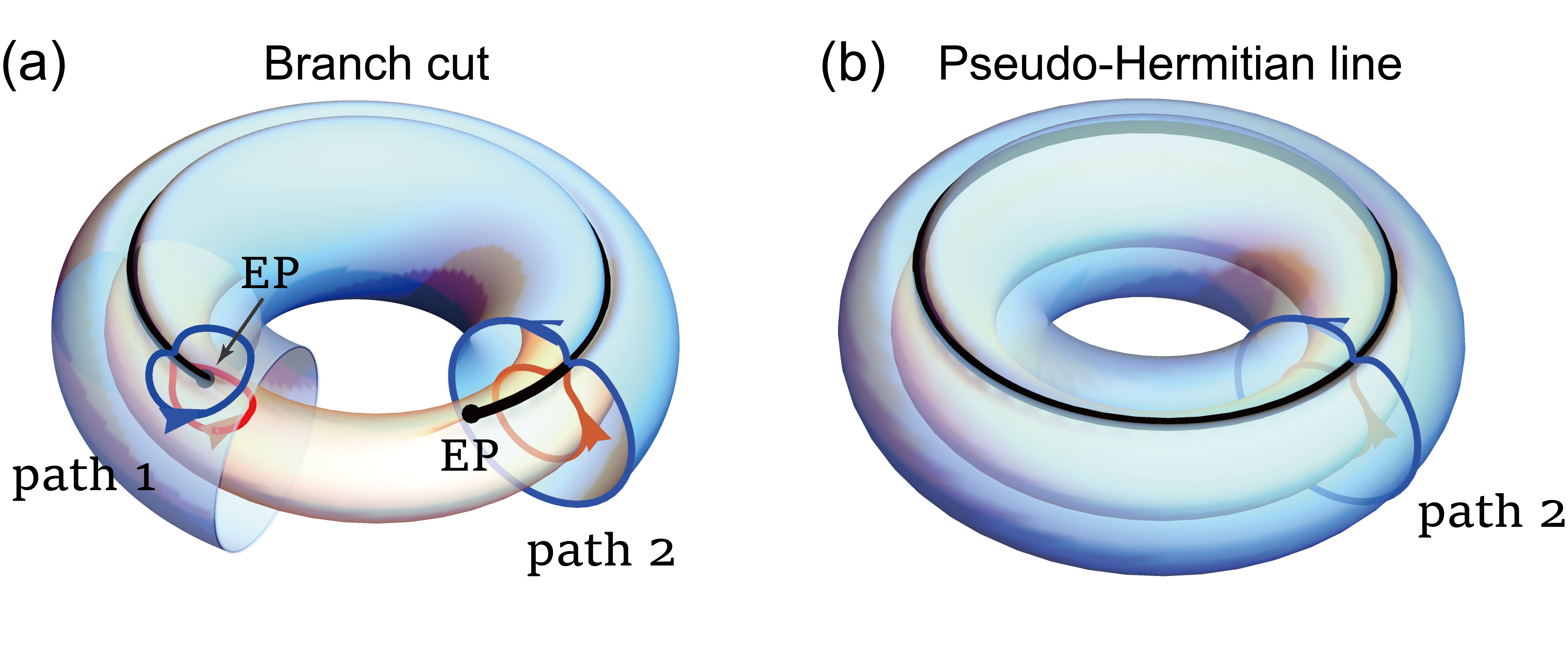}
\caption{Real energy surfaces on a two-dimensional Brillouin zone (torus). (a) Two energy surfaces, inner red and outer blue surfaces, cross along the branch cut (black arc) that ends at two exceptional points (EPs, two black dots). The blue and red portions of loops represent the trajectories for adiabatic parameter changes on the blue and red energy surfaces, respectively. Conventional state exchange occurs along the path encircling an EP, path 1. The state exchange also occurs along path 2, without encircling an EP. (b) The EPs merge and vanish, leaving the pseudo-Hermitian line (black circle). Despite the absence of EPs in the parameter space, state exchange still occurs along path 2.}
\label{fig:fig_EP_PHL}
\end{figure}

We consider generic two-state parametric Hamiltonian, 
\begin{eqnarray}
    H(\alpha) = {\bf d}(\alpha) \cdot \bm \sigma + d_0(\alpha) \sigma_0,
\label{eq:ham22}
\end{eqnarray}
where ${\bf d} = {\bf d}_R + i {\bf d}_I$ with ${\bf d}_R$, ${\bf d}_I \in \mathbb R^3$, $d_0 \in \mathbb C$, $\bm \sigma$ is the vector of Pauli matrices, $\sigma_0$ is the $2\times2$ identity matrix, and $\alpha = \{\alpha_i\}$, $\alpha_i \in \mathbb R$ are parameters of Hamiltonian.
The last term, $d_0 \sigma_0$, just gives a constant shift in complex energy, so we set $d_0 = 0$ for simplicity. The complex-energy spectrum then explicitly reads as
\begin{eqnarray}
    E_\pm = \pm \sqrt{{\bf d}_R^2 - {\bf d}_I^2 + 2i{\bf d}_R \cdot {\bf d}_I}.
\label{eq:ham22_energy}
\end{eqnarray}
If $\bold{d}_R \cdot \bold{d}_I = 0$, the Hamiltonian has real or 
conjugate pairs of pure imaginary eigenenergies depending on the sign of ${\bf d}_R^2 - {\bf d}_I^2$. This condition has been referred to as the pseudo-Hermiticity condition, the existence of invertible $2\times2$ matrix $\eta$ satisfying $\eta H \eta^{-1} = H^\dagger$. Since the co-dimension of pseudo-Hermitian manifolds is one, the parameter sets satisfying the pseudo-Hermiticity condition are points and lines in one and two spatial dimensions, respectively. These are protected under a small perturbation in the Hamiltonian. In this work, we are mainly concerned with two spatial dimensions, so we refer to the pseudo-Hermitian manifold as a PHL. Figure~\ref{fig:fig_EP_PHL} exemplifies the form of eigenenergies of a two-band model in a two-dimensional parameter space with a real PHL (${\bf d}_R^2 < {\bf d}_I^2$) in the cases with and without EPs, respectively. Along the PHL (black curve), the two eigenenergies are conjugate pairs, and the real eigenenergies coincide (see Appendix~\ref{appendix:PHLs}). At EPs, ${\bf d}_R^2 = {\bf d}_I^2$ and the eigenenergies degenerate. For PHLs in multiband case, we use the effective $2\times2$ Hamiltonian describing only two bands among others to validate the above definition.

\subsection{Non-contractible PHLs} Now, we focus on the state exchanges after adiabatic evolution along a loop in the parameter space. In an infinite parameter space such as $\mathbb{R}^2$, a PHL alone does not drive state exchanges. Although states are exchanged as a path crosses a PHL, they always meet an even number of times: the path forms a loop, and the PHL is either an extended line or a closed loop. The loop has to include EPs to have a non-trivial state exchange, and the effects are usually attributed to the EPs. Also, the PHL is non-contractible when EPs are attached to the ends of the PHL or surrounded by the PHL. It has been widely studied that such non-separability of band structures appears when one considers a loop that encircles EPs [path 1 in Fig.~\ref{fig:fig_EP_PHL} {\bf a}]. If a PHL forms a loop without EPs, the PHL can be continuously deformed to annihilate itself (see Appendix~\ref{appendix:cBCs} for a detailed example).

In contrast, the PHL can derive non-trivial effects on state exchanges when the topology of the parameter space changes. Consider a 2D Brillouin zone where the parameters are two momenta $k_x$ and $k_y$. Due to periodic boundary conditions, they form a torus. The Riemann surfaces of complex eigenenergies lie on top of the torus, as shown in Fig.~\ref{fig:fig_EP_PHL}. On a torus, both an encircling loop and PHL can be contractible or non-contractible. If one of the loops is contractible, a PHL alone does not make any non-trivial effects as in an infinite parameter space. Similarly, there is no state exchange when both loops are non-contractible and encircle the same hole in the torus. However, if both loops are non-contractible and encircle different holes of the torus, the PHL and the encircling loop can intersect an odd number of times, resulting in non-trivial effects [Fig.~\ref{fig:fig_EP_PHL}{\bf b}]. Consequently, when a non-contractible PHL exists in the Brillouin zone, the band structure becomes non-separable along a loop that crosses the PHL, as shown in Fig.~\ref{fig:fig_EP_PHL}{\bf b}. Note that in this case, the stability of PHL comes from the periodicity of the parameter space.

In summary, non-trivial state exchanges are directly related to the PHLs and do not depend on whether they are connected to EPs or not. The homotopy classification can be applied to both cases with and without EPs, only by invoking PHLs, for a given encircling loop. Without EPs, the PHLs in an infinite plane do not give rise to a non-trivial effect, but they do in parameter spaces with different topologies, such as torus. This is due to the existence of non-contractible loops in those spaces. Also, non-separable bands can exist without any EPs in the space.

\begin{figure}
\centering
\includegraphics[width=1.0\linewidth]{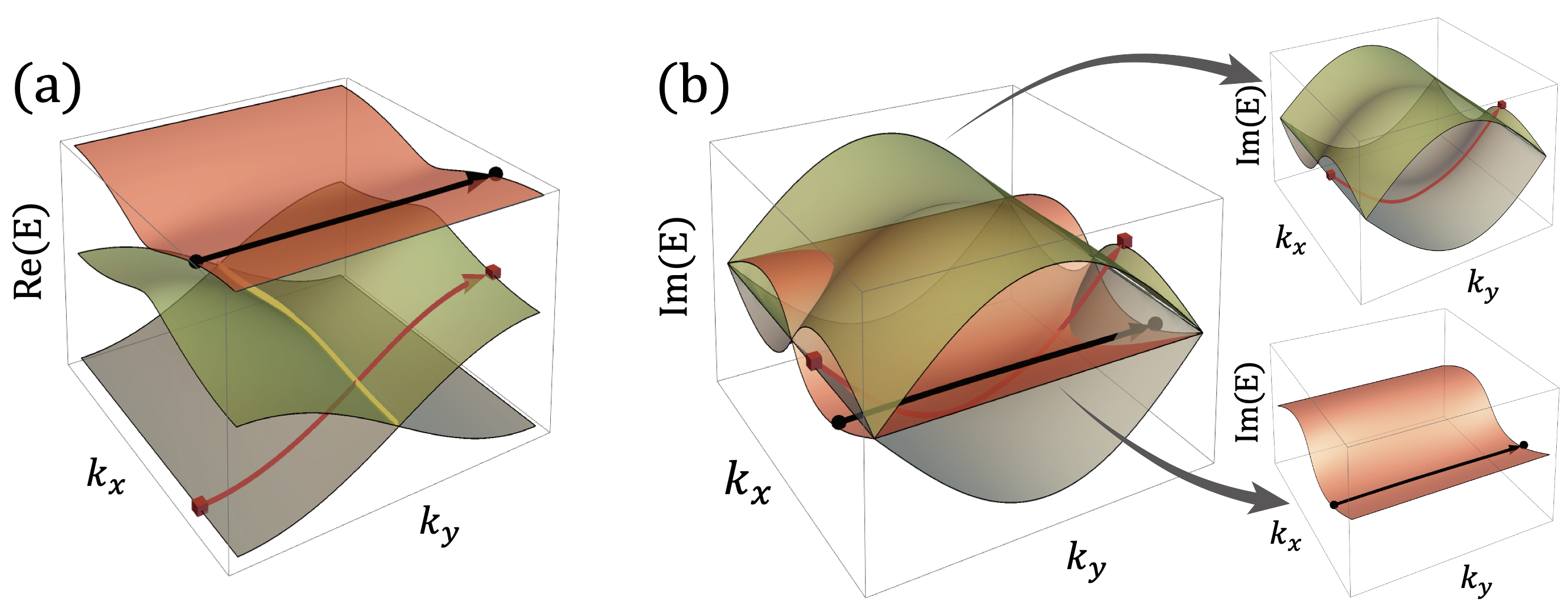}
\caption{Complex eigenenergies with PHLs. (a) Real and (b) imaginary parts of the complex eigenenergies of the Hamiltonian in Eq.~(\ref{eq:ham1}) when $(s_1 , s_2) = (0.5, 0)$. The red, green, and gray bands are the first, second, and third bands, according to the order of real parts of complex eigenenergies. The green and gray bands are not separable and are mapped on Fig.~\ref{fig:fig_EP_PHL}(b). The yellow line represents a PHL. The black spheres and red cubes represent the initial and final states when $k_y$ changes from $-\pi$ to $\pi$. The arrows form closed loops because of the periodicity of the 2D Brillouin zone.}
\label{fig:tBCs}
\end{figure}

\section{PHLs in three-bands Hamiltonains}

\subsection{Non-trivial multibands without EPs} We exemplify the PHL by providing a model Hamiltonian that is given as \cite{Koe23}
\begin{eqnarray}
    H(s_1,s_2) &=& (1- s_1 - s_2 ) h( k_x, k_y ) \nonumber \\
    &&\hspace{30pt}+ s_1 h(\pi, k_y ) + s_2 h(k_x, \pi),
\label{eq:ham1}
\end{eqnarray}
where
\begin{gather}
    h( k_x, k_y ) = \\\nonumber
    \begin{pmatrix}
        1 + 2 \sqrt{2} - e^{i k_x} & -i (1+e^{i k_x}) & 0 \\
        -i (1+e^{i k_x}) & e^{i k_x} - e^{i k_y} & -i (1+e^{i k_y})\\
        0 & -i (1+e^{i k_y}) & -1 - 2 \sqrt{2} + e^{i k_y}
    \end{pmatrix},
\end{gather}
$k_x,k_y$ are crystalline momenta, and $s_1,s_2$ represent external parameters that control the configurations of the EPs and the PHL. 
Figure~\ref{fig:tBCs} shows the real and imaginary band structure of the Hamiltonian when $(s_1, s_2)=(0.5,0)$. In the real band structure, the PHL occurs along the $k_x$ direction (yellow line), indicating that the real eigenenergies of the two bands are degenerate, while the corresponding imaginary eigenvalues are not.

As we consider a non-contractible loop that encircles along the $k_y$ direction (red lines in Fig.~\ref{fig:tBCs}), state exchange occurs after the encircling. The eigenstate switching effect can be formally characterized by the classification of the permutation group \cite{Ryu22}. In this example, the homotopy class corresponds to
$1^1 2^1$, where $C^n$ represents $n$ $C$-cycle bands (see Appendix~\ref{appendix:classification}). The class $1^1 2^1$ represents one 1-cycle separable band (first real) and one 2-cycle non-separable band (second and third real). The cycle $C$ corresponds to the denominator in the fractional winding number of the complex eigenvalues \cite{Lee16}. 
The imaginary bands also have the same state exchange effect. However, the imaginary bands have three PHLs with two additional imaginary band crossings.

A heuristic way to illustrate a non-contractible PHL is to consider zone folding of the Hatano--Nelson model and add perturbations (see Appendix~\ref{appendix:PHL_HN}). In Hermitian systems, zone folding generates band degeneracy, which can be immediately gapped by the addition of infinitesimal perturbations. In contrast, a PHL in a non-Hermitian system cannot be gapped out, and the location of the PHL is shifted in 2D space.

\begin{figure}
    \centering
    \includegraphics[width=1.0\linewidth]{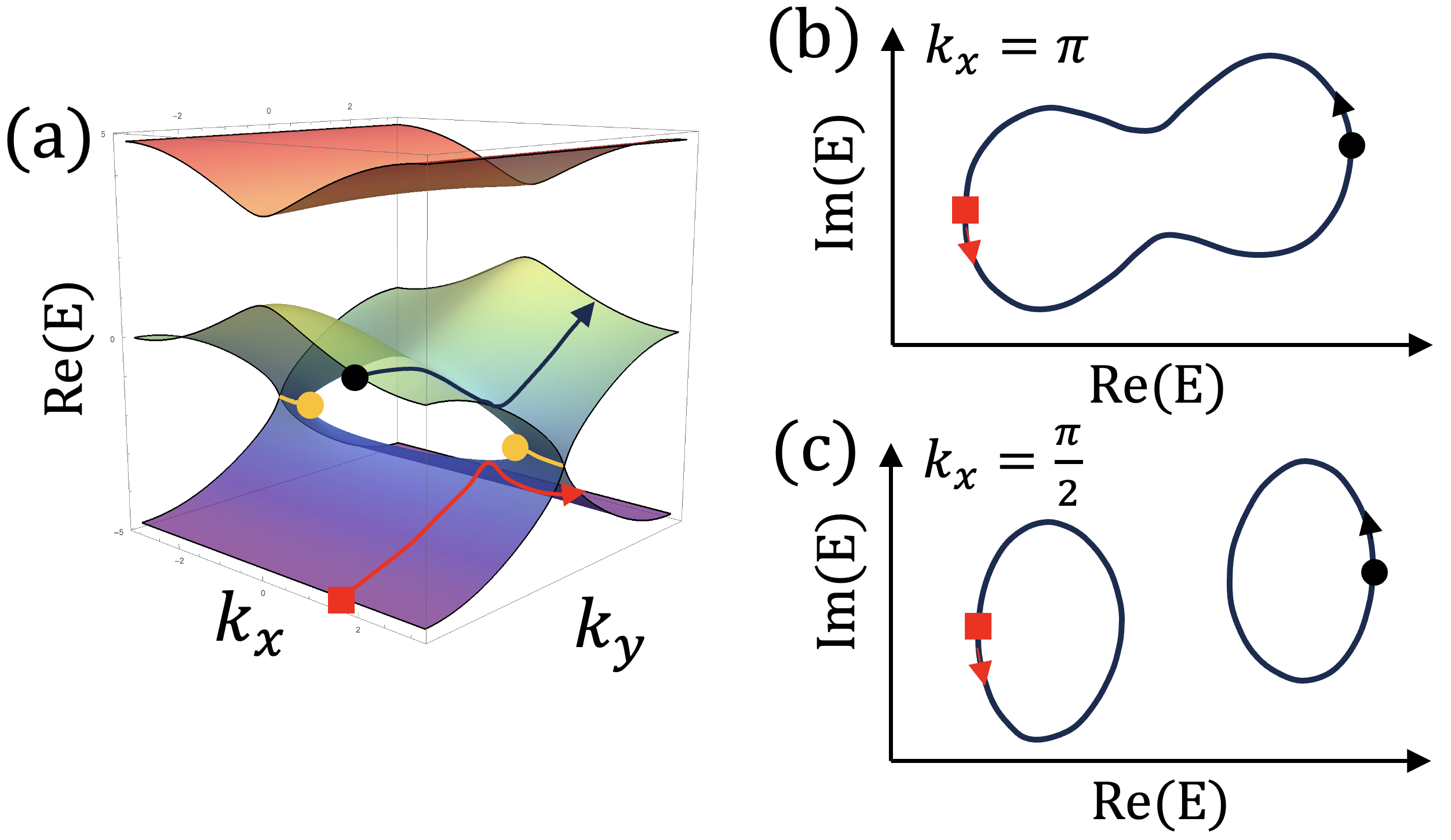}
    \caption{Complex eigenenergies with a pair of EPs. (a) Real parts of the complex eigenenergies of the Hamiltonian in Eq. (3) when $s_2 = 0.3$. The two lower bands are mapped on Fig.~\ref{fig:fig_EP_PHL}(a). The two yellow points and lines represent EPs and corresponding branch cuts. The black circle and red rectangle are initial states when $k_y = - \pi$, and the black and red lines denote closed loops when $k_y$ changes from $-\pi$ to $\pi$ with $k_x = \pi / 2 $. (b) Complex eigenenergies of the two lower bands on the complex plane when $k_x = \pi$. The black circle and red rectangle are initial states when $k_y = - \pi$, and the black lines denote closed loops when $k_y$ changes from $-\pi$ to $\pi$ with $k_x = \pi $. (c) Complex eigenenergies of two lower bands on the complex plane when $k_x = \pi/2$. The black circle and red rectangle correspond to those in (a).}
    \label{fig:transitionEPs}
\end{figure}

\subsection{Phase transitions via EPs} While non-contractible PHLs are protected against small perturbations, when a sufficiently large perturbation is added, the generation of EPs can change the complex energy structure where the PHL emerges or terminates. 
The band structures of corresponding 2D systems can be represented by the homotopy classes of state changes for non-contractible encircling loops. Accordingly, a Lifshitz transition of the band structure occurs by the creation and annihilation of EPs. Figure~\ref{fig:transitionEPs} shows the energy spectrum of the Hamiltonian in Eq.~(\ref{eq:ham1}) with $s_2 =0.3$. A pair of EPs between the two lowest bands emerges. If we consider a loop that encircles along the $k_y$-direction with $k_x=\pi$, a state exchange occurs, which corresponds to the $1^{1}{2}^1$ class [Fig.~\ref{fig:transitionEPs}{\bf b}]. On the other hand, if we consider a loop at $k_x=\pi /2$, the homotopy class of the state exchange changes to $1^3$ [Fig.~\ref{fig:transitionEPs}{\bf c}]. Thus, there exist two kinds of encircling along the $k_y$ direction, having $1^12^1$ and $1^3$ classes. This is different from the $s_2=0$ case where there is only the $1^12^1$ class. As $s_2$ increases, the EPs move to the zone boundary and finally disappear when they meet. Then the state exchange has the $1^3$ class only, and the two bands become isolated or line-gapped \cite{She18, Kaw19b}. Note that non-contractible loops along $k_x$ are always in the class $1^3$. PHLs can also be terminated by exceptional lines (see Appendix~\ref{appendix:PHL_HN}); real and imaginary PHLs change into imaginary and real PHLs, respectively, via exceptional lines. Next, we consider EPs and PHLs in multiband systems.

\begin{figure}
    \centering
    \includegraphics[width=1.0\linewidth]{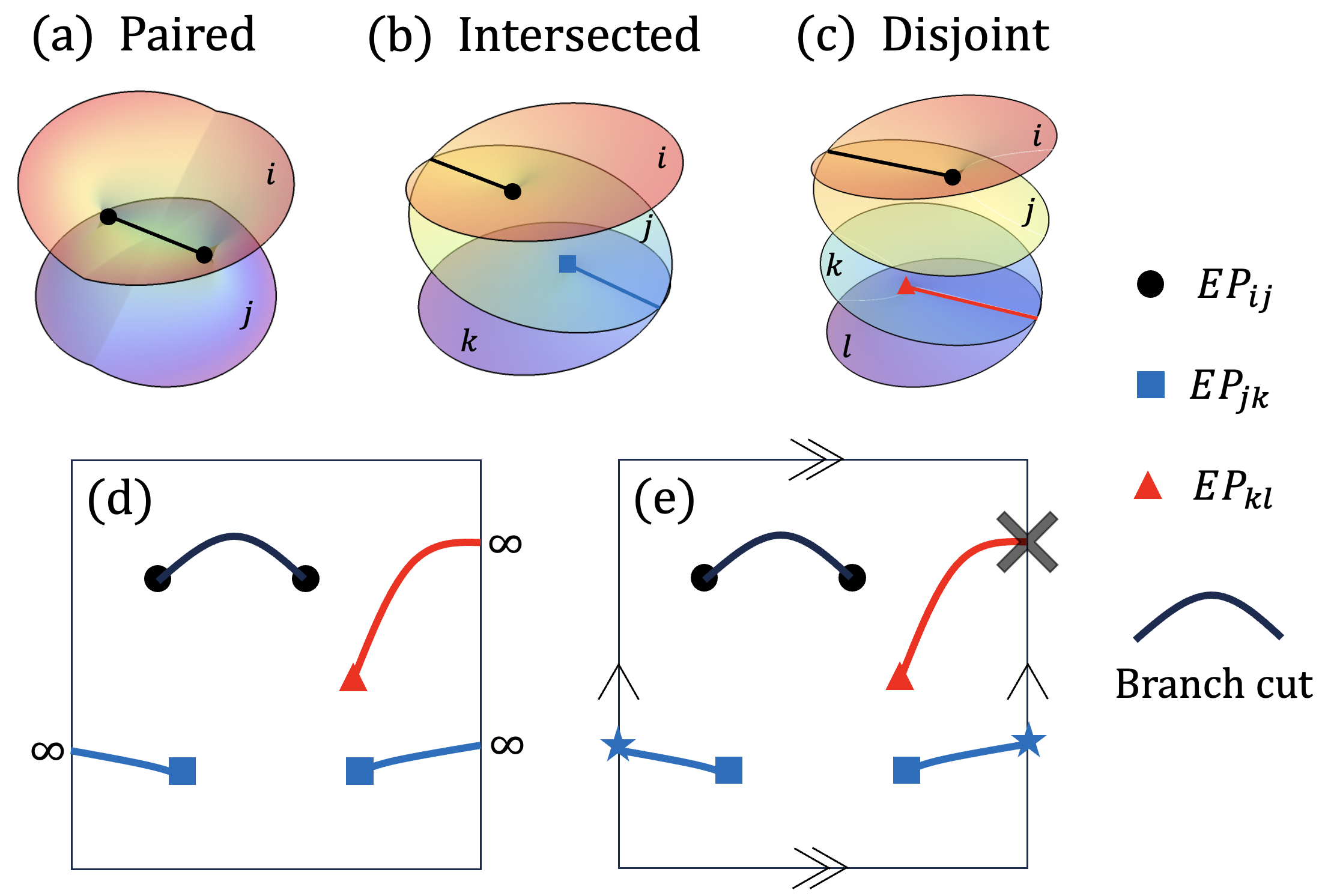}
    \caption{Three kinds of two EPs. (a) Paired EPs with a shared branch cut, (b) intersected EPs, and (c) disjointed EPs with independent branch cuts. Combinations of EPs in (d) 2D infinite and (e) periodic space. Black and blue represent paired EPs, and red represents an unpaired EP. Unpaired EPs (a single EP, or intersected or disjointed EPs) cannot exist in (e) 2D periodic space because of the compactness of the space.}
    \label{fig:twoeps}
\end{figure}

\section{Generalization of PHLs}

\subsection{Generalization of multibands and multiple EPs} We now generalize to the case of multiband systems with more than two bands. We first consider a second-order EP with square root branch behavior, $f(z) = z^{1/2}$, and branch cuts that are PHLs between pairs of EPs or an EP and infinity. When more than two bands are considered, we can label each energy band at a given momentum $\mathbf{k}$ such that $\textrm{Re} \epsilon_{1}(\mathbf{k})\ge \textrm{Re} \epsilon_{2}(\mathbf{k}) \ge ... \ge \textrm{Re} \epsilon_{N}(\mathbf{k}) $. Accordingly, we can label the EP associated with the $i$ and $j$ states as $EP_{ij}$. The two distinct EPs can be categorized as having a paired, intersected, or disjointed relation (Fig.~\ref{fig:twoeps}). Paired EPs are two EPs with the same associated states, $i$ and $j$. The branch cut connects the two EPs or extends to infinity in parameter space. Intersected EPs are two EPs that share only one state $j$, e.g., $EP_{ij}$ and $EP_{jk}$. In this case, the corresponding branch cut from two EPs extends to infinity in parameter space. Disjointed EPs correspond to two EPs without any shared states, where branch cuts can cross each other. Regardless of the relation, branch cuts originating from EPs create arcs between paired EPs or between an EP and infinity in a given parameter space. As a result, more than two EPs can always be decomposed into combinations of (i) an EP with a branch cut extended to infinity and (ii) paired EPs with a branch cut between the EPs. Both intersected and disjointed EPs can be considered as combinations of single EPs with branch cuts extended to infinity in the parameter space.

\subsection{Multiple EPs in different 2D spaces} The periodicity of the parameter space gives constraints on the relations of the EPs. We first consider infinite space. A 2D infinite parameter space is mapped onto a Riemann sphere with one point at infinity, and we consider various kinds of EPs located on the sphere. First, an EP is located on the sphere and a branch cut from the EP should be extended to the point at infinity [red triangle and line in Fig.~\ref{fig:twoeps}{\bf d}]. Next, paired EPs share a branch cut between them or have two branch cuts that extend to the point at infinity, depending on the selection of branch cuts. As a result, any combination of EPs can exist in 2D parameter spaces since there is no constraint. An example is the quasi-bound states in deformed microcavities in 2D parameter space. All possible classes have been realized in deformed microcavities exhibiting four quasi-bound states \cite{Ryu22}. A 2D space with one-dimensional periodicity that is mapped into a cylinder is the same \cite{Ryu23}.

On the other hand, we next consider a 2D Brillouin zone that is mapped into a torus that contains neither infinity nor boundary, i.e., a compact and closed surface. A single EP cannot exist in the space alone since a branch cut from the EP cannot be extended to infinity [red triangle and line in Fig.~\ref{fig:twoeps}{\bf e}]. Only paired EPs with a shared branch cut can exist in the space [black circles, blue rectangles, and lines in Fig.~\ref{fig:twoeps}{\bf e}] \cite{Yan21}, while unpaired EPs (a single EP, or intersected or disjointed EPs) cannot exist because of the compactness of the space. Multiple paired second-order EPs are topologically equivalent to higher-order EPs, which can exist in a 2D Brillouin zone \cite{Koe23}.

\begin{figure}[t]
    \centering
    \includegraphics[width=1.0\linewidth]{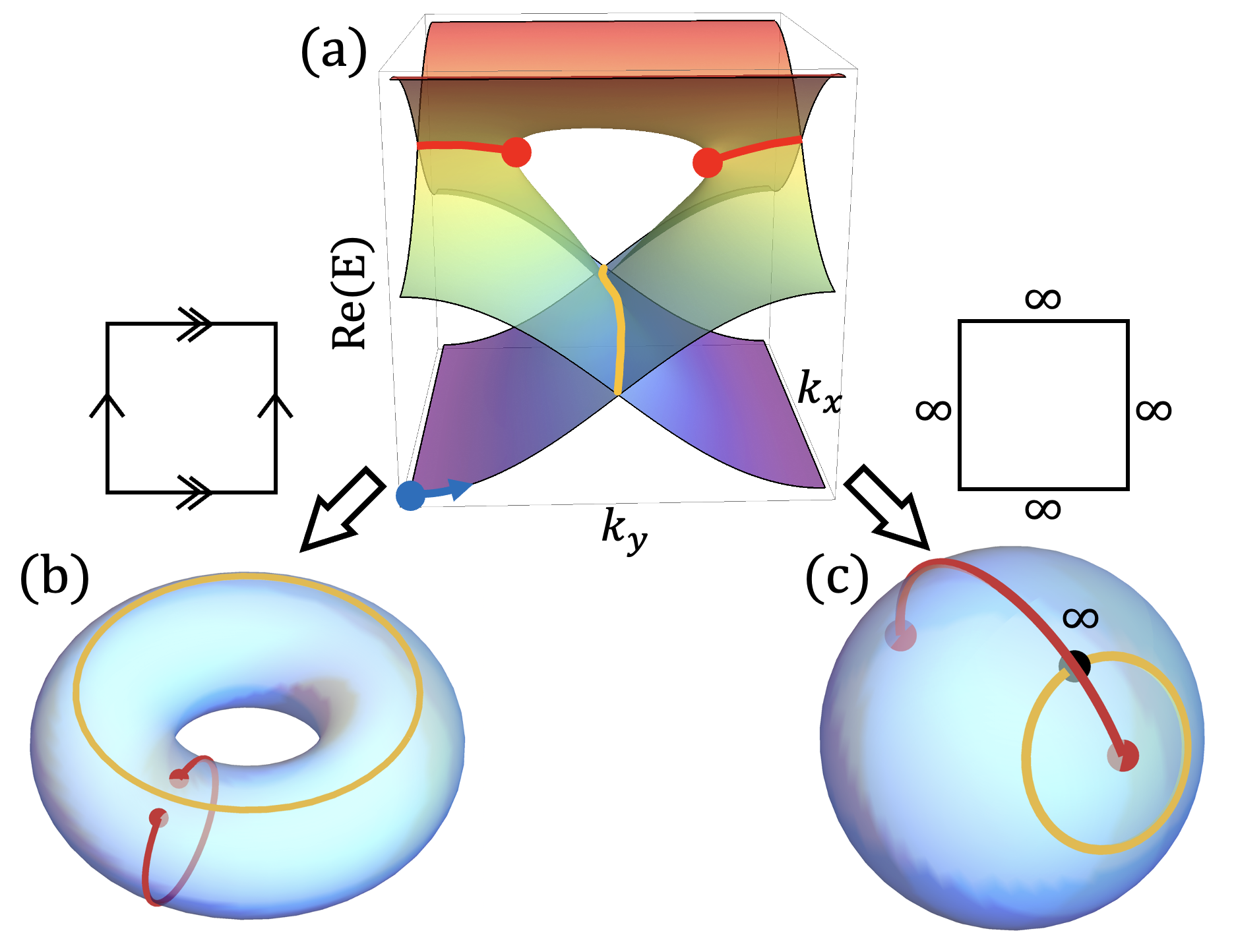}
    \caption{Complex eigenenergies in two different given spaces. (a) Real parts of the complex eigenenergies of the Hamiltonian in Eq. (3) when $(s_1, s_2) = (0.25, 0)$, comprising a PHL (orange line) and two EPs (red circles) with two associated branch cuts (red lines). An initial state (blue circle) returns to its origin after a three-cycle rotation on the Brillouin zone boundary. (b) A PHL, pair of EPs, and associated branch cuts on a torus as a 2D periodic space. (c) A PHL, pair of EPs, and associated branch cuts on a Riemann sphere with one point at infinity (violet) as a 2D infinite space. The PHL is a non-contractible loop due to the presence of an EP in the loop, while the line is contractible if there is no EP in the loop.}
    \label{fig:combined}
\end{figure}

\subsection{Non-contractible PHLs due to the given space and EPs.}
An EP can give rise to the stability of an intersected PHL. Figure~\ref{fig:combined}{\bf a} shows the real energy spectrum of the Hamiltonian $H$ in Eq.~(\ref{eq:ham1}) with $(s_1, s_2) = (0.25,0.0)$. Two EPs exist between the first and second real bands, and a PHL appears between the second and third real bands. The band structures can be represented by the state exchanges along two non-contractible loops, namely along $k_x$ and $k_y$, which are both in classes $1^3$ or $1^12^1$. Note that when combining the two loops, first along $k_x$ and then along $k_y$, the state exchange is in class $3^1$.
With this setup we consider the topological stability of a PHL in the presence of EPs. If the parameter space is a Brillouin zone forming a torus, the PHL is stable due to the topology of the torus, regardless of EPs [Fig.~\ref{fig:combined}{\bf b}]. For an infinite space, the PHL cannot be removed without passing through EPs. Note that the stability is not due to the topology of the parameter space but to the existence of the EP that the PHL encircles. This is explicitly shown in a Riemann sphere formed by infinite space with a point of infinity, as illustrated in Fig.~\ref{fig:combined}{\bf c}. In this case, the PHL is a loop starting at infinity and encircles one of the EPs. The PHL is stable since it is non-contractible due to the encircled EP.

\begin{figure*}[t]
    \centering
    \includegraphics[width=1.0\linewidth]{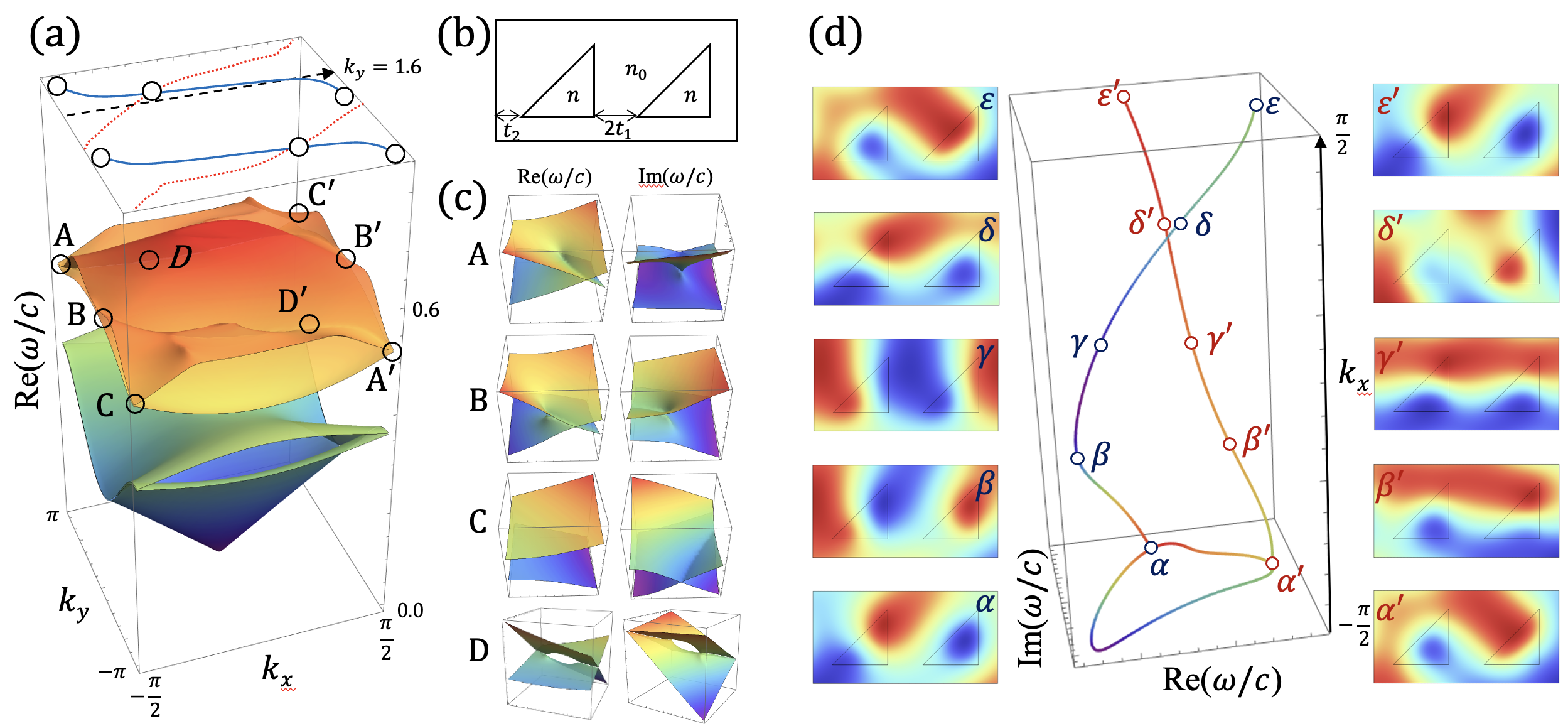}
    \caption{Four bands in a photonic crystal and braids via PHLs. (a) The lowest four energy bands in (b) a photonic crystal made of two lossy triangular cavities with a complex refractive index $n$. In the upper plane, the blue solid and red dotted lines represent simplified real and imaginary PHLs, respectively. The circles indicate the presence of EPs in tiny regions. These regions approach zero in size, meaning pairs of EPs come closer and finally merge into vortex points as the difference between $t_1$ and $t_2$ decreases. (c) The enlarged complex energy bands in the regions where there are EPs or pairs of EPs. (d) The complex energy braid of third and fourth bands when $k_x$ changes from $-\pi/2$ to $\pi/2$ ($\alpha$, $\alpha '$: $k_x = -\pi/2$, $\beta$, $\beta '$: $k_x = -\pi/4$, $\gamma$, $\gamma '$: $k_x = 0$, $\delta$, $\delta '$: $k_x = \pi/4$, $\varepsilon$, $\varepsilon '$: $k_x = \pi/2$) when $k_y = 1.6$, along with the corresponding wavefunctions. For the visibility of the braids, we colored the curves with respect to the imaginary value of the energy. The first and the third columns show the wavefunction amplitudes at denoted points.}
    \label{fig:braid}
\end{figure*}

\section{Realization of non-contractible PHLs in photonic crystals}

\textcolor{black}{Multiband topology in non-Hermitian tight-binding models with complicated hoppings, such as non-reciprocal, long-range, and complex-valued hoppings, has been realized in recent experiments. In a system of optical ring resonators, a lattice model can be realized along a frequency synthetic dimension by simultaneously modulating the phase and amplitude of the resonators, enabling complex hoppings \cite{Wan21c, Wan21b}. Additionally, the non-Hermitian lattice models with non-reciprocity, such as the Hatano-Nelson models (see Appendix~\ref{appendix:PHL_HN}) and non-Hermitian Su-Schrieffer-Heeger models, have also been experimentally realized in various experiments, e.g. electric circuits \cite{Hel20}, mechanical systems \cite{Gha20}, and acoustic cavities \cite{Zha23b}. Since these experiments can exactly reproduce the theoretical results of tight-binding models, they can be expected to implement our model Hamiltonians intact.}

As another possible physical realization of our models, we consider two-dimensional photonic crystals with lossy cavities. The cavity losses introduce effective non-reciprocity, enabling the study of nontrivial point-gap topology, the non-Hermitian skin effect, and multiple EPs \cite{Zho21, Zho23}. Here, we study PHLs and state exchanges in a photonic crystal. Consider a lattice of triangular cavities with two bases in a unit cell, as shown in Fig. \ref{fig:braid}{\bf b}. The positions of two triangular cavities are represented by $t_1=t_0-\delta$ and $t_2=t_0+\delta$ with asymmetry $\delta=0.01$ (in a unit of lattice constant). The refractive index of cavities is $n=2-0.8i$, where the complex number implies the loss. We solve Maxwell's equations for this system using COMSOL MULTIPHYSICS, employing finite-element methods. The real eigenvalues of the lowest four bands are shown in Fig. \ref{fig:braid}{\bf a}, where we focus on the third and fourth lowest bands. Four pairs of EPs are identified: one in each region $A$ ($A'$) and $B$ ($B'$), and two in region $D$ ($D'$) as shown in Fig. \ref{fig:braid}{\bf a}. Real (imaginary) PHLs connect these EPs, depicted as blue-solid (red-dotted) lines in the figure. The complex energy structures in small regions $A$ to $D$ are shown in Fig. \ref{fig:braid}{\bf c}. After the adiabatic evolution along the loop parallel to $k_x$ with $k_y=1.6$, the states in the third and fourth bands exchange states, resulting in a non-trivial effect (black dashed line in Fig \ref{fig:braid}{\bf a}). This state exchange is also shown in Fig. \ref{fig:braid}{\bf d} with wavefunction amplitude of the states within the unit cell. Two initial states are represented by blue and red circles when $k_x = -\pi/2$ (see $\alpha$ and $\alpha '$ in Fig.~\ref{fig:braid}{\bf d}). As the $k_x$ increases, two states follow the different curves passing through blue and red circles indicated by the Greeks and thus the curves generate braids. Finally, two states exchange each other when $k_x = \pi/2$ (see $\varepsilon$ and $\varepsilon '$ in Fig.~\ref{fig:braid}{\bf d}). As $\delta$ approaches zero, the EPs in the region $D$, and those in $A$ and $A'$ ($B$ and $B'$) merge into the vortex points \cite{She18}. However, the PHLs still remain, and thus, the topological structures of photonic multibands determined by PHLs.

\section{Discussion}

We have demonstrated that complex eigenenergies and non-orthogonal eigenstates in non-Hermitian systems manifest unique topological phenomena, in stark contrast to their Hermitian counterparts. We have also unveiled the role of pseudo-Hermitian lines in dictating the topology of non-Hermitian systems. We show that the PHL plays a prominent role in driving non-trivial state exchanges along a loop in the parameter space. In 2D lattice systems, the PHL can create non-trivial state exchanges without any EPs in the Brillouin zone. Also, the band structures can be represented by the classes of state exchange along non-contractible loops. The PHL can be stable owing to not only EPs but also the topology of the parameter space. \textcolor{black}{We have discussed multiband structures in a photonic crystal with lossy materials as physical systems to implement state exchanges as well as PHLs.} We expect this work to shed light on versatile applications and realizations of non-trivial, non-Hermitian systems without the need to consider exceptional points that generally require experimental fine-tuning. For instance, utilizing further-neighbor hopping and symmetry-breaking perturbations, the fractional winding numbers of complex energies can be systematically realized in controllable experiments \cite{Wan21b, Pat22, Yu22b, Zha23b}.

Recently, we became aware of a related recent work of Ref~\cite{Leh24}. There, the authors show that the nodal spectral function is protected by the braiding properties of bands in a one-dimensional Brillouin zone, which is consistent with our result of stability of PHLs. Note that the pseudo-Hermitian point plays the same role in the homotopy classification once we select the one-dimensional loops in the two-dimensional systems.

\section*{Acknowledgements}

We acknowledge financial support from the Institute for Basic Science in the Republic of Korea through the project IBS-R024-D1.

\appendix
\section*{Appendix}

\section{Pseudo-Hermitian lines in a two-band model}
\label{appendix:PHLs}

\begin{figure}[b!]
\centering
\includegraphics[width=\figsizeone\linewidth]{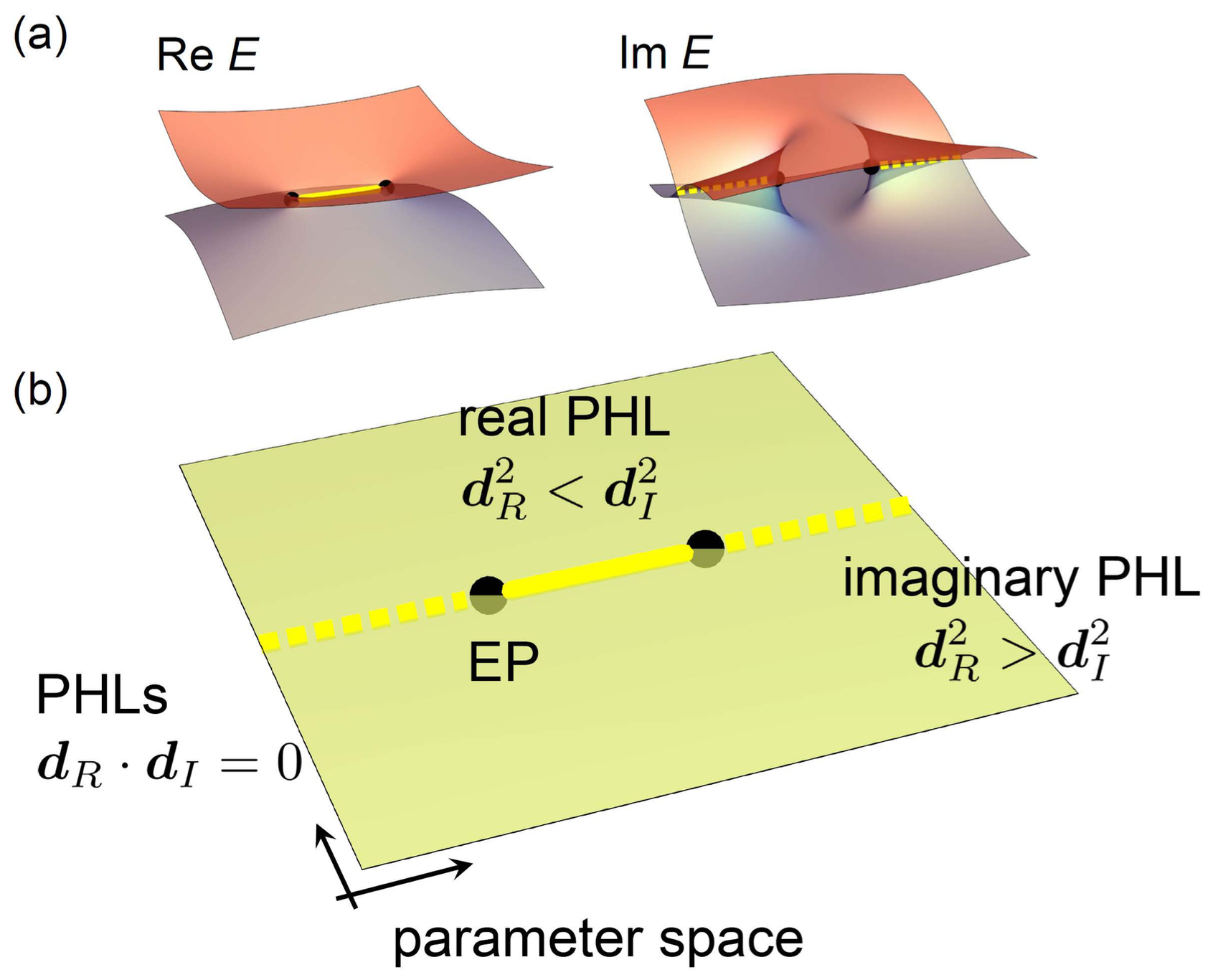}
\caption{Schematic figure of pseudo-Hermitian lines (PHLs) in parameter space. (a) Real and imaginary eigenenergies of a two-band model in parameter space. The yellow solid (dashed) line is a real (imaginary) PHL where the real (imaginary) energies coincide. (b) Real and imaginary PHLs in parameter space. Exceptional points (black spheres) are located at the points where the real and imaginary PHLs meet.}
\label{fig:fig_PHL}
\end{figure}

Figure~\ref{fig:fig_PHL} exemplifies the eigenenergies of a two-band model. Figure~\ref{fig:fig_PHL}(a) shows real and imaginary eigenenergies on the parameter space. On the pseudo-Hermitian lines (PHLs), yellow solid and dashed lines in Fig.~\ref{fig:fig_PHL}, the condition ${\bf d}_R \cdot {\bf d}_I=0$ is satisfied. They form lines in the parameter space [Fig. \ref{fig:fig_PHL}(b)]. For ${\bf d}_R^2 < {\bf d}_I^2$, the two eigenenergies are conjugate pairs and the real eigenenergies coincide. We denote this PHL as real PHL (yellow solid lines in Fig. \ref{fig:fig_PHL}). For ${\bf d}_R^2 > {\bf d}_I^2$, eigenenergies are real and the imaginary eigenenergies coincide, denoted as the imaginary PHL (yellow dashed lines in Fig.~\ref{fig:fig_PHL}). When the real and imaginary PHLs meet, the condition ${\bf d}_R^2 = {\bf d}_I^2$ is satisfied and the points are exceptional points (EPs).

\section{Removable pseudo-Hermitian lines}
\label{appendix:cBCs}

\begin{figure}[t]
    \centering
    \includegraphics[width=\figsizeone\linewidth]{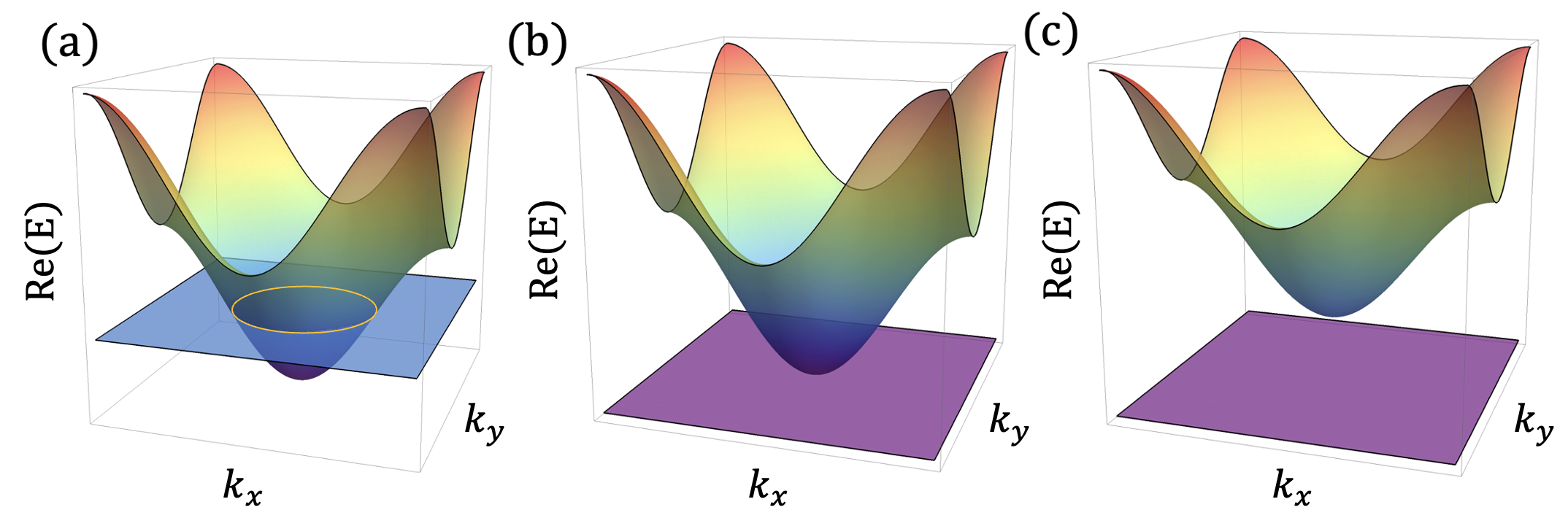}
    \caption{PHLs can be removed by continuous transformation. Two bands of the Hamiltonian, Eq.~(\ref{eq:hamA2}), when (a) $\alpha = 1$, (b) $\alpha = 2$, and (c) $\alpha = 3$. The orange ring represents a pseudo-Hermitian ring.}
    \label{fig:tPHL}
\end{figure}

An example of a non-singular band touching appears in the bilayer square lattice model \cite{Rhi19}. The modified non-Hermitian Hamiltonian is given by
\begin{gather}
\label{eq:hamA2}
    H = 
    \begin{pmatrix}
        - \cos{k_x} - \cos{k_y} - i & \alpha - \cos{k_x} - \cos{k_y} - i\\
        \alpha - \cos{k_x} - \cos{k_y} - i &  - \cos{k_x} - \cos{k_y} - i
    \end{pmatrix},
\end{gather}
which has two eigenenergies $E_1 (k) = - \alpha$ and $E_2 (k) = \alpha - 2 \cos{k_x} - 2 \cos{k_y} - 2 i$. Two bands are shown in Fig.~\ref{fig:tPHL} depending on $\alpha$. The pseudo-Hermitian ring [Fig.~\ref{fig:tPHL}(a)] can be removed via trivial real band touching [Fig.~\ref{fig:tPHL}(b)] with different imaginary band energies.

\section{Classification of eigenstate switching}
\label{appendix:classification}

Here, we review the classification of eigenstate exchanges after the adiabatic encirclings in the parameter space of non-Hermitian systems \cite{Ryu22, Ryu23}. In Hermitian systems, a state with a gap remains its instantaneous eigenstate for adiabatic change of parameters. In non-Hermitian systems, however, states can be changed even for cyclic processes. This is due to the existence of non-separable bands, which originated from EPs or PHLs. The state exchange after the encircling in parameter space can be described by holonomy or permutation matrices. If one is not interested in the details of the transformation during the encircling, the overall effects are just the permutation of states. Thus, one can classify the effect of state exchange based on the permutation group~\cite{Ryu22}. Conjugacy classes of permutation group of $N$ objects are denoted by
\begin{eqnarray}
    (1^{\alpha_1} 2^{\alpha_2} 3^{\alpha_3} \cdots N^{\alpha_N}),
\end{eqnarray}
with a constraint $\sum_{n=1}^N n \alpha_n = N$, $\alpha_i$ are non-negative integer. The notation represents that there are $n_i$ of $i$-cycles with $i=1, 2, \cdots, N$. The same notation can be applied to the classes of state exchanges. For example, there are three classes for three band systems, which are denoted by
\begin{eqnarray}
    (1^3), \ (1^1 2^1), \ (3^1).
\end{eqnarray}
Thus, there are classes in which three bands are separable, two bands are non-separable, and all three bands are non-separable.

\section{PHLs derived from Brillouin zone folding and multiband topology in two-dimensional Hatano--Nelson models}
\label{appendix:PHL_HN}

\subsection{Hatano--Nelson model}
\begin{figure}[t]
\centering
\includegraphics[width=\figsizeone\linewidth]{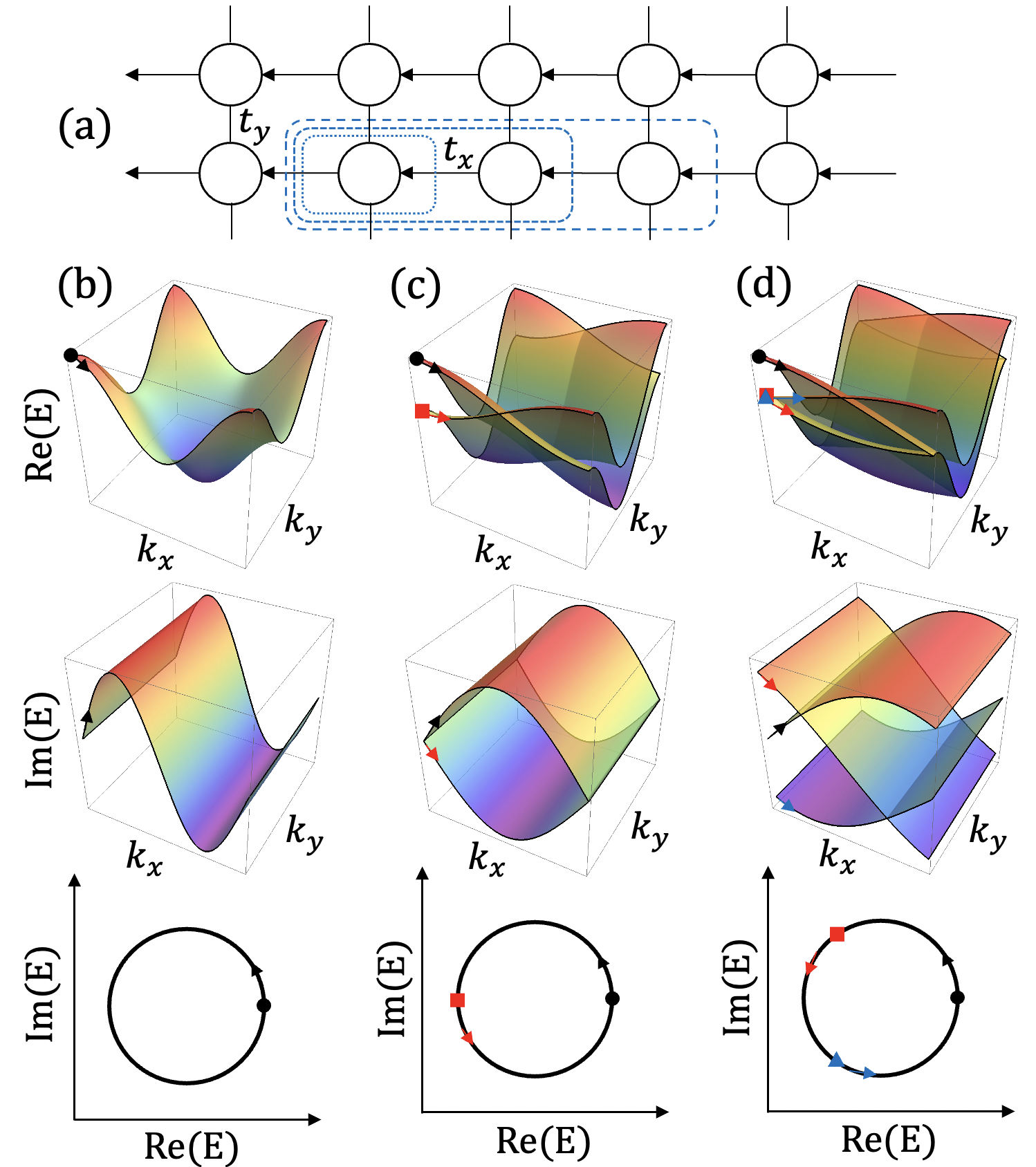}
\caption{(a) Two-dimensional lattice model, namely a Hatano--Nelson model with unidirectional hopping $t_x$ for the $x$ axis and a normal chain with hopping $t_y$ for the $y$ axis. Dotted, dashed, and long-dashed boxes respectively represent unit cells with one, two, and three sites. Real parts, imaginary parts, and adiabatic changes of complex eigenenergies of Hamiltonians of which unit cells have (b) one, (c) two, and (d) three sites. The black circles, red rectangles, and blue triangles represent the initial and final states when $k_x$ changes from $0$ to $2\pi$.}
\label{fig:BC_HN}
\end{figure}

To understand non-separable bands in multiband non-Hermitian systems from a different perspective, we consider a well-known two-dimensional (2D) lattice model, the Hatano--Nelson model with unidirectional hoppings for the $x$ axis and a normal chain for the $y$ axis [Fig.~\ref{fig:BC_HN}(a)]. The 2D Hamiltonian is
\begin{gather}
    H = \sum_{x=1}^{x=\infty} \sum_{y=1}^{y=\infty} [ t_x c_x^\dagger c_{x+1} + t_y c_{y+1}^\dagger c_y + t_y c_y^\dagger c_{y+1}] ,
\label{eq:ham_HN}
\end{gather}
where the eigenenergies are given by
\begin{gather}
    E = t_x e^{i k_x} + 2 t_y \cos{k_y} .
\label{eq:ham_HN_energy}
\end{gather}
As $k_x$ increases from $0$ to $2 \pi$ when $k_y = 0$, the complex eigenenergy has a point gap and the winding number equals $1$ since the phase of the complex eigenenergy around an arbitrary reference point inside a complex energy loop accumulates $2 \pi$ during the eigenenergy return to the initial eigenenergy [Fig.~\ref{fig:BC_HN}(b)]. The non-zero winding number represents the non-Hermitian skin effect, which is the same as a one-dimensional Hatano--Nelson model where all eigenstates are localized on a boundary under open boundary conditions.

\begin{figure}[t]
\centering
\includegraphics[width=\figsizeone\linewidth]{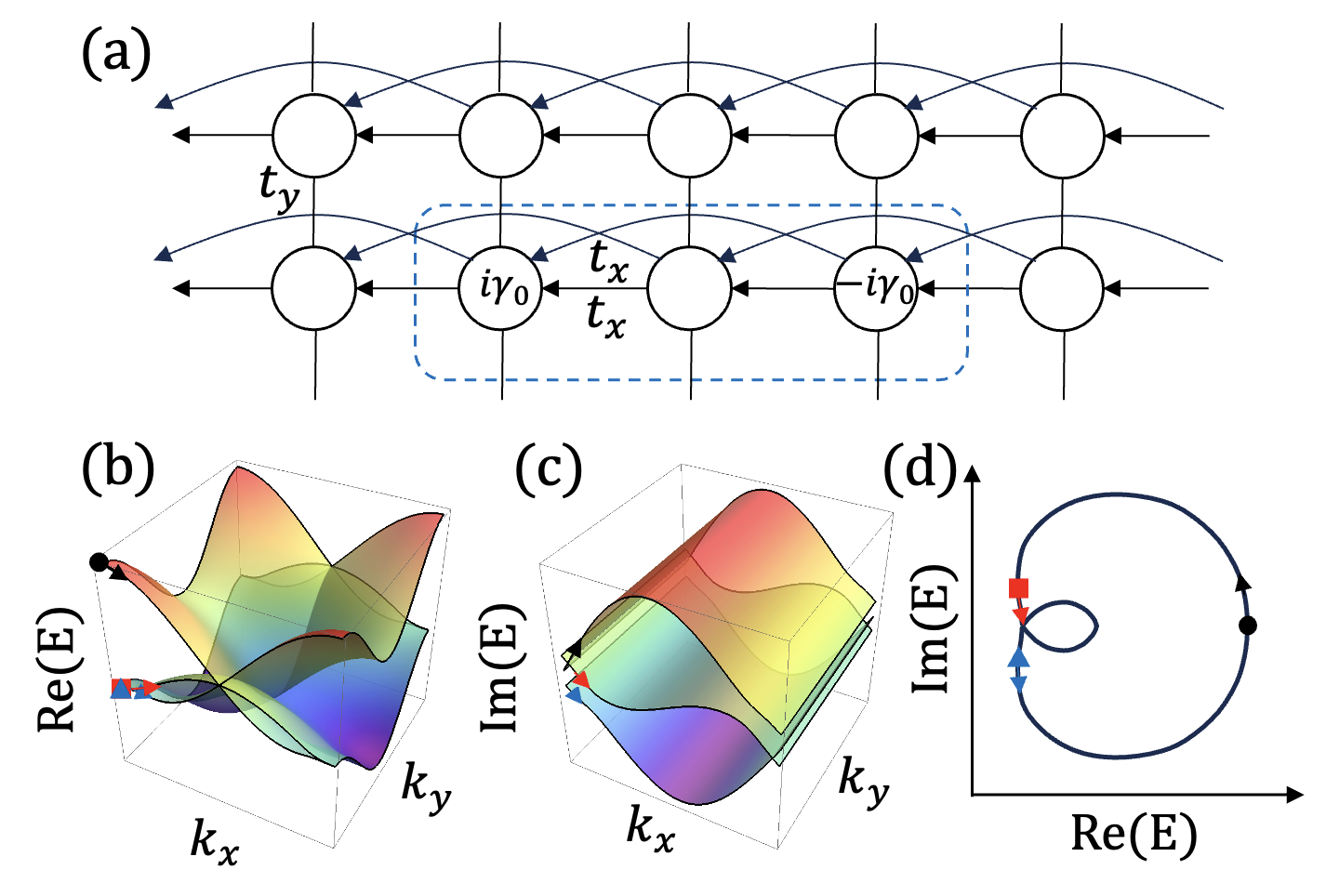}
\caption{(a) Two-dimensional three-band lattice model, namely a Hatano--Nelson model with nearest and next nearest unidirectional hopping $t_x$ for the $x$ axis and a normal chain with hopping $t_y$ for the $y$ axis. The long-dashed box represents the unit cell with three sites. (b) Real parts, (c) imaginary parts, and (d) adiabatic changes of complex eigenenergies of the Hamiltonian, Eq.~(\ref{eq:ham_frac}). The black circles, red rectangles, and blue triangles represent the initial and final states when $k_x$ changes from $0$ to $2\pi$.}
\label{fig:fractional_winding}
\end{figure}

\subsection{PHLs and fractional winding numbers in multiband non-Hermitian systems}
By redefining the unit cells to contain two sites, a two-band model is constructed by Brillouin zone folding. The two-band Hamiltonian is
\begin{gather}
\label{eq:ham_22}
    H( k_x, k_y ) = 
    \begin{pmatrix}
        2 t_y \cos{k_y} & t_x \\
        t_x e^{i k_x} & 2 t_y \cos{k_y}
    \end{pmatrix} ,
\end{gather}
where the eigenenergies are given by
\begin{gather}
    E = \pm t_x e^{\frac{i k_x}{2}} + 2 t_y \cos{k_y} .
\label{eq:ham_22_energy}
\end{gather}
Here, the two energy bands do not have degenerate points in the Brillouin zone, and the lines of intersection between them are PHLs. There is a real PHL at $k_x = \pi$ and an imaginary PHL at $k_x = 0$. As $k_x$ increases from $0$ to $2 \pi$ when $k_y = 0$, the two bands exchange with each other after one period of $k_x$ [Fig.~\ref{fig:BC_HN}(c)]. After two periods of $k_x$ (from $0$ to $4 \pi$), since a band returns to its initial band, the winding number is $1/2$, which is a fractional winding number in non-Hermitian multiband systems. We note that the PHLs do not disappear but instead shift their positions in 2D space under small perturbations, unlike the nodal lines in Hermitian systems.

Redefining the unit cells to contain three sites, the three-band model is
\begin{gather}
    H( k_x, k_y ) = \\\nonumber 
    \begin{pmatrix}
        2 t_y \cos{k_y} & t_x & 0 \\
        0 & 2 t_y \cos{k_y} & t_x \\
        t_x e^{i k_x} & 0 & 2 t_y \cos{k_y}
    \end{pmatrix} ,    
\end{gather}
where there are three bands and two real PHLs at $k_x = 0$ and $k_x = \pi$ and two imaginary PHLs at $k_x = \pi / 2$ and $k_x = 3 \pi /2$ [Fig.~\ref{fig:BC_HN}(d)]. The winding number is $1/3$ since a band returns to its initial band after three periods of $k_x$ (from $0$ to $6 \pi$).

We add next nearest neighbor hopping to the three-band model and thus the Hamiltonian is
\begin{gather}
\label{eq:ham_frac}
    H( k_x, k_y ) = \\\nonumber 
    \begin{pmatrix}
        2 t_y \cos{k_y} + i \gamma_0 & t_x & t_x \\
        t_x e^{i k_x} & 2 t_y \cos{k_y} & t_x \\
        t_x e^{i k_x} & t_x e^{i k_x} & 2 t_y \cos{k_y} - i \gamma_0 
    \end{pmatrix} ,
\end{gather}
where there are three bands along with two real PHLs and two imaginary PHLs when $\gamma_0 = 0.5$ (Fig.~\ref{fig:fractional_winding}). The fractional winding number $W/C$ equals $2/3$ since the initial band returns to its initial band after three periods of $k_x$ (from $0$ to $6 \pi$) and the phase of complex eigenenergy around an arbitrary reference point inside the complex energy loop accumulates $4 \pi$ during three periods of $k_x$. The numerator $W$ and denominator $C$ of the fractional winding number are topological invariants from the non-Hermitian skin effect and the homotopy equivalence class of the multiband, respectively.

\subsection{PHLs are terminated by non-Hermitian degeneracy}
\begin{figure}[t]
\centering
\includegraphics[width=\figsizeone\linewidth]{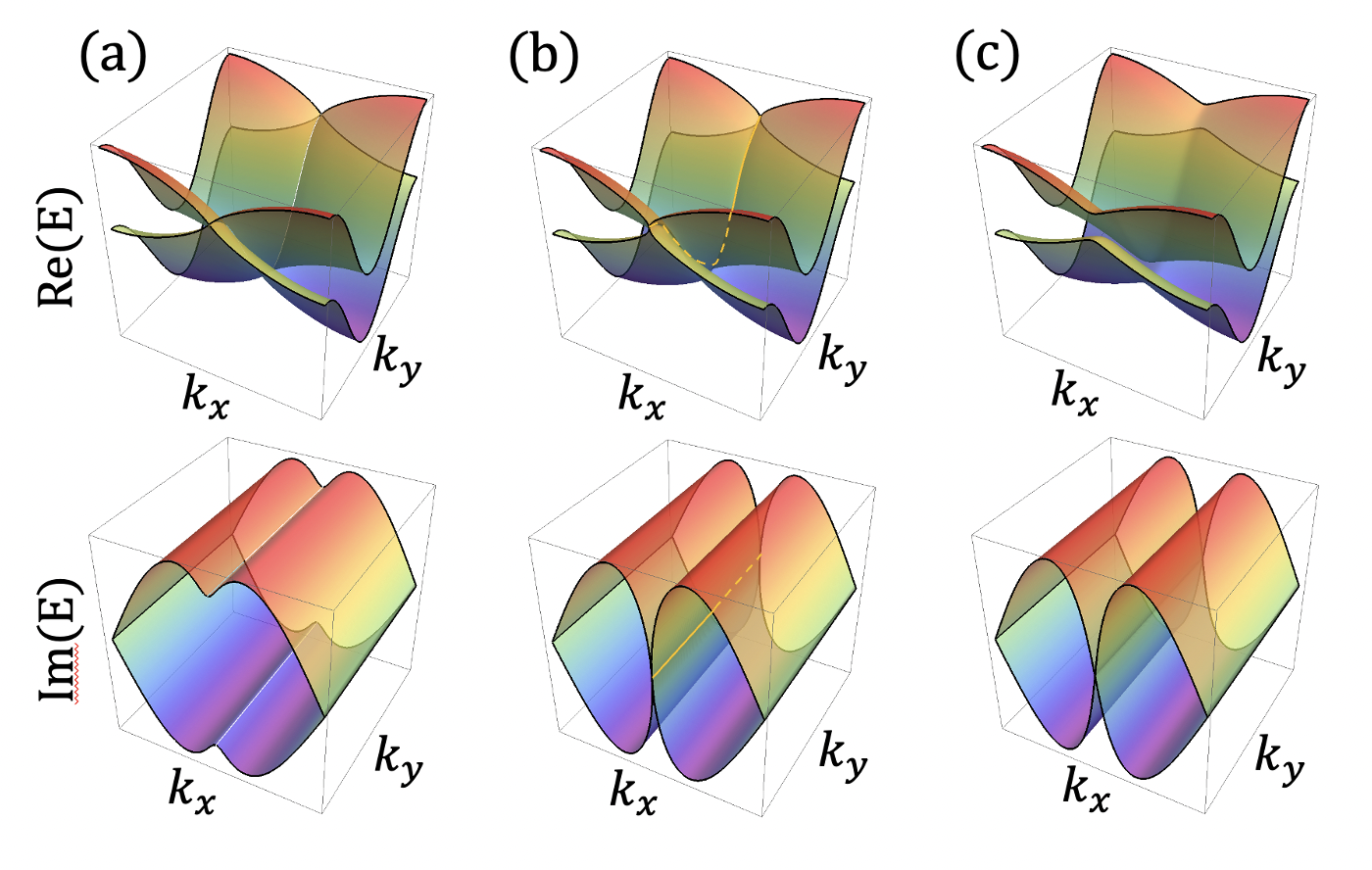}
\caption{Real and imaginary parts of complex eigenenergies of the Hamiltonian Eq.~(\ref{eq:ham_22}) with onsite perturbations, (a) $\epsilon_0 = 0.9$, (b) $\epsilon_0 = \epsilon_{EL} = 1.0$, and (c) $\epsilon_0 = 1.1$. There is an exceptional line [yellow lines in (b)] where real and imaginary PHLs merge at $k_x=\pi$ when $\epsilon = \epsilon_{EL}$ .}
\label{fig:BC_ELs}
\end{figure}

Pseudo-Hermitian lines are topologically protected in 2D space. They cannot be removed or added by continuous transformations when the PHLs are non-contractible loops in the 2D Brillouin zone, i.e., a torus (Figs.~\ref{fig:BC_HN} and \ref{fig:fractional_winding}), while they can be removed in 2D infinite parameter spaces. As onsite energy $\pm \epsilon_0$ in the Hamiltonian in Eq.~(\ref{eq:ham_22}) increases, while the real PHL does not change, the imaginary parts of two bands at $k_x = \pi$ become closer [Fig.~\ref{fig:BC_ELs}(a)] and finally make an imaginary PHL at $\epsilon_0 = \epsilon_{EL}$ [Fig.~\ref{fig:BC_ELs}(b)]. At $\epsilon_0 = \epsilon_{EL}$, real and imaginary PHLs merge into an exceptional line. As $\epsilon_0$ increases further, the real PHL changes into the imaginary PHL [Fig.~\ref{fig:BC_ELs}~(c)].

\begin{figure}[b]
\centering
\includegraphics[width=\figsizeone\linewidth]{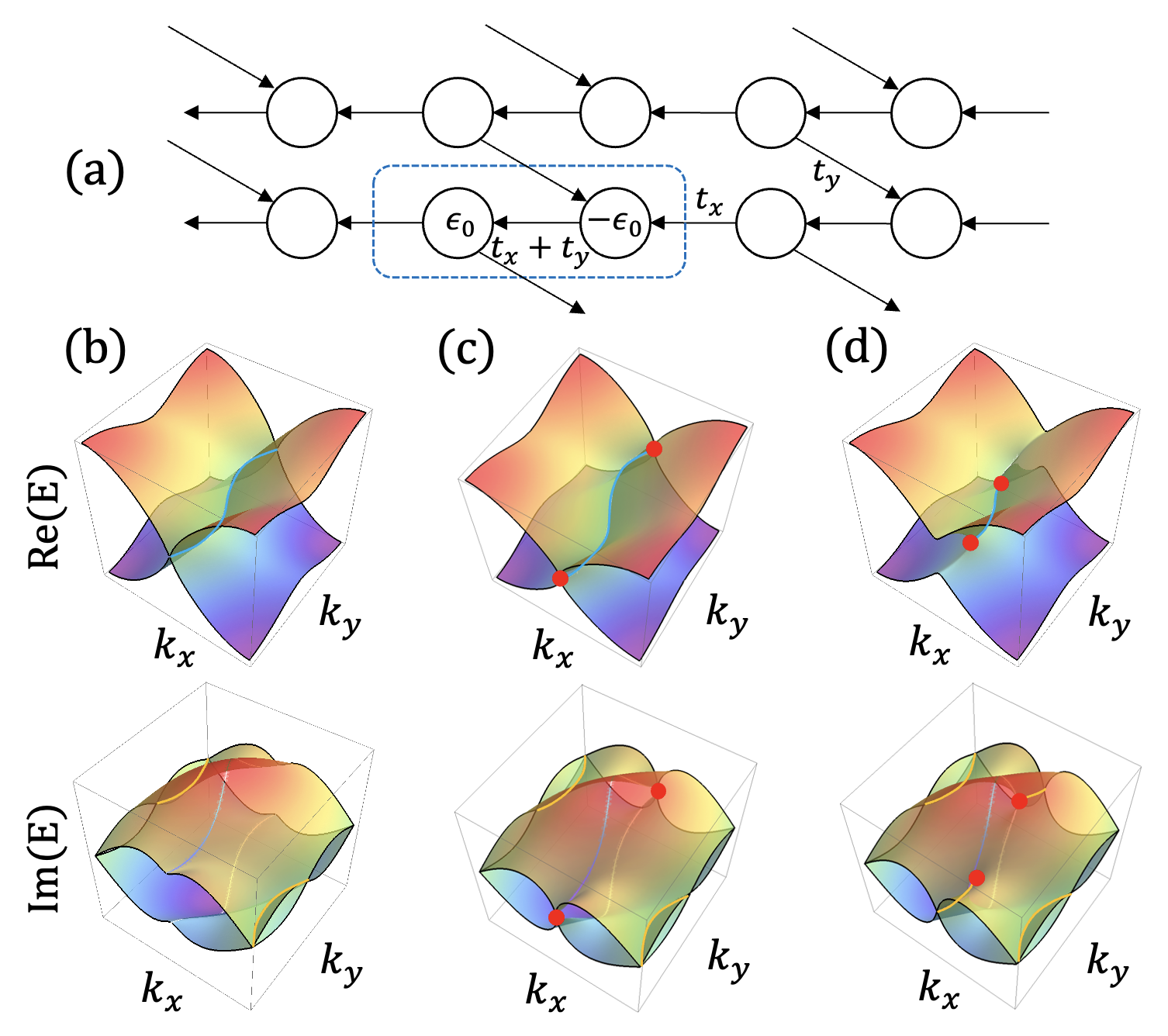}
\caption{(a) Two-dimensional lattice model with alternative types of unidirectional hopping. The dashed box represents a unit cell with two sites. Real and imaginary parts of complex eigenenergies of the Hamiltonian Eq.~(\ref{eq:ham_22_2D}) with onsite perturbations: $\epsilon_0 = 0.8$ (b), $\epsilon_0 = \epsilon_{EP} \sim 0.866$ (c), and $\epsilon_0 = 1.0$ (d). The red circles and blue and orange lines represent a pair of EPs and branch cuts between the EPs.}
\label{fig:BC_EPs}
\end{figure}

The next example is the case where EPs terminate PHLs. Consider a 2D lattice model that is a two-band model with alternative types of unidirectional hopping [Fig.~\ref{fig:BC_EPs}(a)]. The two-band Hamiltonian is
\begin{gather}
\label{eq:ham_22_2D}
    H( k_x, k_y ) = 
    \begin{pmatrix}
        \epsilon_0 & t_x + t_y \\
        t_x e^{i k_x} + t_y e^{i k_y} & -\epsilon_0
    \end{pmatrix} .
\end{gather}
There are real and an imaginary PHLs when $\epsilon_0 = 0.8$ [Fig.~\ref{fig:BC_EPs}(b)]. EPs emerge at $(k_x , k_y) = (\pi, 0)$ on the real PHL when $\epsilon_0 = \epsilon_{EP}$, and thus the real PHL becomes a branch cut between EPs as branch points of the Riemann surface of complex energy multibands [Fig.~\ref{fig:BC_EPs}(c)]. As $\epsilon_0$ increases, an imaginary branch cut emerges and then there are two EPs and corresponding real and imaginary branch cuts between the two EPs [Fig.~\ref{fig:BC_EPs}(d)]. As $\epsilon$ increases further, the real branch cut disappears when the two EPs merge and disappear. It is noted that topologically protected non-contractible PHLs can be terminated and removed by non-Hermitian degeneracy such as EPs and exceptional lines.

\end{document}